\documentclass[fleqn,12pt,twoside]{article}
\usepackage{espcrc1}
\usepackage{epsf}

\usepackage{graphicx}

\newcommand{\AmS}{{\protect\the\textfont2
  A\kern-.1667em\lower.5ex\hbox{M}\kern-.125emS}}

\begin{document}



\begin{center}
{\Large Generalized Parton Distributions~: Experimental aspects} 
\vskip 1. cm
{\bf M. Guidal}\\ 
\vskip 1. cm
{\it IPN Orsay, F-91406 Orsay, France}\\
\end{center}

\section{INTRODUCTION}

Much of the internal structure of the nucleon has been revealed
during the last two decades through the \underline{inclusive}
scattering of high energy leptons on the nucleon in the
Bjorken -or ``Deep Inelastic Scattering" (DIS)- regime 
($Q^2,\nu\gg$ and $x_B=\frac{Q^2}{2M\nu}$ finite). 
Simple theoretical 
interpretations of the experimental results and quantitative
conclusions can be reached in the framework of QCD, when 
one sums over all the possible hadronic final states. For 
instance, {\it unpolarized} DIS
brought us evidence of the quark and gluon substructure 
of the nucleon, quarks carrying about 45\% of the nucleon
momentum. Furthermore, {\it polarized} DIS revealed that no more than 
about 25\% of the spin of the nucleon is carried by the quarks' 
intrinsic spin.

Now, with the advent of the new generation of high-energy,
high-luminosity lepton accelerators combined with large
acceptance spectrometers, a wide variety of 
\underline{exclusive} processes in the Bjorken regime can 
be envisaged to become accessible experimentally. Until
recently, no sound theoretical formalism was available for
a systematic interpretation, in particular for the electroproduction
of photons and mesons. A unified description is now under way through the
formalism of new ``Generalized Parton Distributions" (GPDs) -also
called ``Skewed Parton Distributions"-.
 
These distributions parametrize the complex structure of the nucleon 
and allow to describe various exclusive processes such as
Virtual Compton Scattering (\cite{Ji97,Rady})
and (longitudinal) vector and pseudo-scalar meson electroproduction 
\cite{Collins97}. The GPDs contain information on the correlations 
between quarks (i.e. non-diagonal elements) and on their transverse 
momentum dependence in the nucleon. As a direct effect of these features, 
Ji also showed~\cite{Ji97} that the second moment of these GPDs gives access 
to the sum of the quark spin and the quark orbital angular momentum to 
the nucleon spin, which may shed light on the so-called nucleon ``spin-puzzle".
Most of these informations are not contained in the 
traditional inclusive parton distributions extracted from inclusive DIS 
which allows to access only partons densities, i.e. diagonal elements.

In this paper, after briefly outlining the formalism of the GPDs in
section~\ref{sec1}, we will discuss some general considerations for 
their experimental study in section~\ref{sec2}~: in particular, the 
relation between GPDs and experimental observables and the need and 
requirements for a dedicated experimental facility. Finally, in
section~\ref{sec3}, we will review and comment the first experimental 
signatures of this physics recently observed by the HERMES and CLAS 
(at JLab) collaborations.

\section{FORMALISM}
\label{sec1}

\begin{figure}[ht]
\epsfxsize=11 cm
\epsfysize=6. cm
\centerline{\epsffile{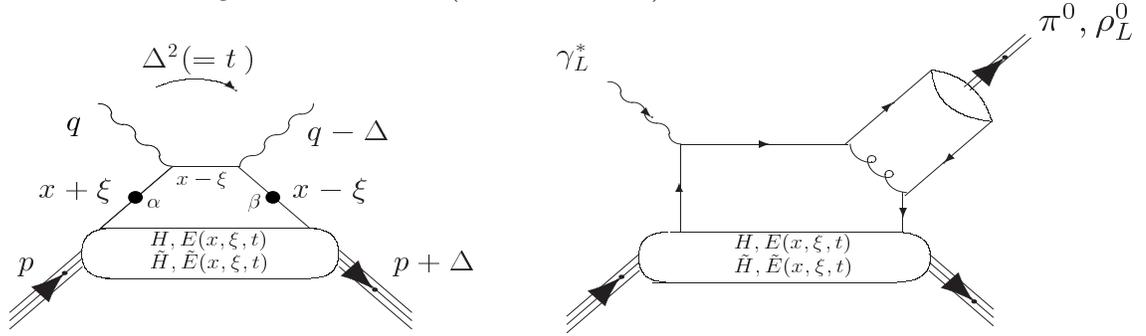}}
\vspace{-1.5cm}
\caption{``Handbag" diagrams~: a) for DVCS (left) and b) for meson 
production (right).}
\label{fig:handbags}
\end{figure}

A few years ago, Ji \cite{Ji97} and Radyushkin \cite{Rady}
have shown that the leading order pQCD amplitude for
Deeply Virtual Compton Scattering (DVCS) in
the forward direction can be factorized in a hard scattering part 
(exactly calculable in pQCD) and a nonperturbative nucleon 
structure part as is illustrated in Fig.(\ref{fig:handbags}-a). 
In these so-called ``handbag" diagrams, 
the lower blob which represents the structure of the nucleon 
can be parametrized, at leading order pQCD, in 
terms of 4 generalized structure functions, the GPDs. These are 
traditionnally called $H, \tilde H, E, \tilde E$, and  
depend upon three variables~: $x$, $\xi$ and $t$. $x-\xi$ is the longitudinal
momentum fraction carried by the initial quark struck by the virtual photon. 
Similarly, $x+\xi$ relates to the final quark going back in the nucleon
after radiating a photon. $-2\xi$ is therefore the longitudinal
momentum difference between the initial and final quarks. In comparison 
to $-2\xi$ which refers to {\it longitudinal}
degrees of freedom, $t$, the standard squared 4-momentum transfer between 
the final nucleon and the initial one, contains {\it transverse} degrees 
of freedom (so-called ``$k_\perp$") as well.

$H$ and $E$ are spin independent, and are often called {\it unpolarized}
GPDs, whereas $\tilde H$ and $\tilde E$ are spin dependent, and are
often called {\it polarized} GPDs. The GPDs $H$ and $\tilde H$ are  
actually a generalization of the parton distributions 
measured in deep inelastic scattering. Indeed, in the forward 
direction, $H$ reduces to the quark distribution and $\tilde H$ to the 
quark helicity distribution measured in deep inelastic scattering. 
Furthermore, at finite momentum transfer, there are
model independent sum rules which relate 
the first moments of these GPDs to the elastic form factors.

The GPDs reflect the structure of the nucleon independently of
the reaction which probes the  nucleon. They can also be accessed 
through the hard exclusive electroproduction of mesons 
-$\pi^0$, $\rho^0$, $\omega$, $\phi$, etc.- 
(see Fig.~(\ref{fig:handbags}-b)) for which a QCD factorization
proof was given recently \cite{Collins97}. According to 
Ref.\cite{Collins97}, the factorization applies when the virtual photon is 
longitudinally polarized because in this case, the end-point contributions
in the meson wave function are power suppressed. 
It is shown in Ref.\cite{Collins97} that the cross section for 
a transversely polarized photon is suppressed by 1/$Q^2$ compared to 
a longitudinally polarized photon. 
Because the transition at the  upper vertices of Fig.~(\ref{fig:handbags}-b) 
will be dominantly helicity conserving 
at high energy and in the forward direction, 
this means that the vector meson should also be predominantly longitudinally 
polarized (notation $\rho^0_L, \omega_L, \phi_L$) for a longitudinal 
photon at QCD leading order and leading twist.

It was also shown in \cite{Collins97} that leading order pQCD predicts
that the vector meson channels ($\rho^0_L$, $\omega_L$, $\phi_L$)
are sensitive only to the unpolarized GPDs ($H$ and $E$) whereas
the pseudo-scalar channels ($\pi^0, \eta,...$) are sensitive only to the 
polarized GPDs ($\tilde{H}$ and $\tilde{E}$). In comparison to meson 
electroproduction, DVCS depends at the same time on {\it both} 
the polarized and unpolarized GPDs.

Another feature to mention, proper to these handbags diagrams, is 
the notion of {\it scaling}. It is predicted that, when asymptotia in $Q^2$
is reached, the differential cross section $\frac{d\sigma}{dt}$
of these ``handbag" mechanisms should show a $\frac{1}{Q^4}$ behavior for
DVCS and a $\frac{1}{Q^6}$ behavior for meson production. These $Q^2$ 
dependences are strong experimental signatures that the appropriate kinematical
regime is reached and are necessary to observe before tempting
to interpret data in terms of GPDs. It has been recently an intense effort 
from the theoretical community to control the corrections (Next to Leading
Order, higher twists, ....) to this scaling behavior~\cite{diehl}. 

\section{GENERAL CONSIDERATIONS FOR AN EXPERIMENTAL STUDY OF THE GPDs}
\label{sec2}

\subsection{Deconvolution issues}

As mentionned in the previous section, the GPDs depend on three 
variables~: $x$, $\xi$ and $t$.
However, it has to be realized that only two of these three variables
are accessible experimentally, i.e. $\xi$ (=$\frac{x_B}{2-x_B}$, fully
defined by detecting the scattered lepton) and $t$ (=$\Delta^2$, see
Fig~(\ref{fig:handbags}-a), fully defined by detecting either the recoil
proton or the outgoing photon or meson). $x$ however is a  
variable which is integrated over, due to the loop in the ``handbag" 
diagrams (see Fig.~(\ref{fig:handbags})).
This means that in general a differential cross section will be
proportional to~:
$\mid \int_{-1}^{+1}d x {H(x,\xi,t) \over {x - \xi + i \epsilon}}+...\mid^2$
(where ``..." stands for similar terms for $E$, $\tilde{H}$, $\tilde{E}$
and ${1} \over {x - \xi + i \epsilon}$ being the propagator of the quark 
between the incoming virtual photon and the outgoing photon -or meson-, see 
Fig.~(\ref{fig:handbags})).
In general, one therefore will measure integrals (with a propagator as 
a weighting function) of GPDs. 

\begin{figure}[htb]
\epsfxsize=8. cm
\epsfysize=8. cm
\centerline{\epsffile{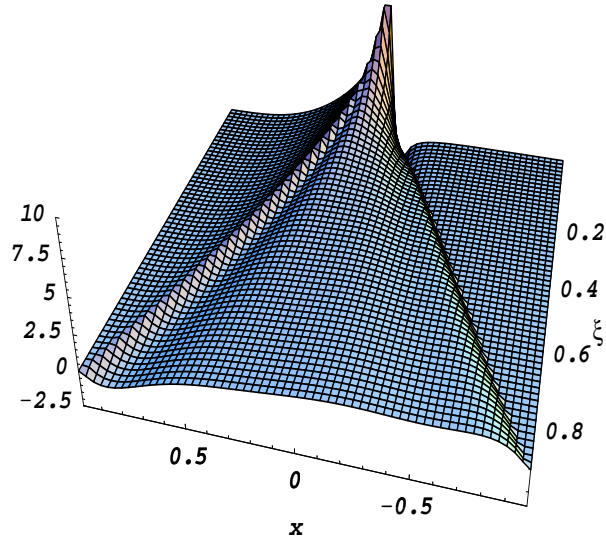}}
\vspace{-1.5cm}
\caption{One model for the GPD $H$ as a function of $x$ and $\xi$ for $t$=0. 
One recognizes for $\xi$=0 the typical shape of a parton distribution (with 
the sea quarks rising as $x$ goes to 0, the negative $x$ part being interpreted 
as the antiquark contribution). Figure taken from~\cite{goeke}.}
\label{fig:spd}
\end{figure}
 
To illustrate this point, Fig.~(\ref{fig:spd}) shows one particular model 
for the GPD $H$ as a function of $x$ and $\xi$ (at $t=0$). One recognizes for
$\xi=0$ a standard quark density distribution with the rise
around $x=0$ corresponding to the diverging sea contribution.
The negative $x$ part corresponds to antiquarks. One sees that the
evolution with $\xi$ is not trivial and that measuring the integral over $x$
of a GPD at constant $\xi$ will not uniquely define it.

A particular exception is when one measures an observable 
proportional to the {\it imaginary} part of the amplitude (for instance, the
beam asymmetry in DVCS which is non-zero in leading order due to the
interference with the Bethe-Heitler process, see section~\ref{sec3}). Then, 
because $\int_{-1}^{+1}d x {H(x,\xi,t) \over {x - \xi + i \epsilon}}=
PP(\int_{-1}^{+1}d x {H(x,\xi,t) \over {x - \xi}})-i\pi H(\xi,\xi,t)$, one 
actually measures directly the GPDs at some specific point, $x=\xi$
(i.e., $H(\xi,\xi,t)$).

For mesons, transverse target polarization observables are also sensitive to a 
different combination of the GPDs, i.e. combinations of the type~: $\int_{-1}^{+1}d x {H(x,\xi,t) \over {x - \xi}}\times
E(\xi,\xi,t)$ (the exact formula is more complicated, see for 
instance~\cite{goeke,polar}). One sees that such transverse spin asymmetries are
sensitive to a {\it product} of the GPDs instead of a sum of their
squares as is the case for a typical differential cross section.

It will therefore be a non-trivial (though a priori not impossible) task 
to actually extract the GPDs from the experimental observables as, to summarize, 
one actually only accesses in general (weighted) integrals of GPDs or 
GPDs at some very specific points or product of these two. In absence of 
any model-independent ``deconvolution" procedure at this moment, one will therefore have to rely on some global model fitting procedure.

It should also be added that GPDs are defined for one quark flavor 
$q$ (i.e. $H^q$, $E^q$,...) similar to standard quark distributions. This ``flavor" separation will require the measurement of several isospin channels ;
for instance, $\rho^0$ production is proportional to (in a succinct notation)
$2/3 H^u + 1/3 H^d$ while $\omega$ production is proportional to 
$2/3 H^u - 1/3 H^d$. Similar arguments apply for the polarized GPDs with
the $\pi^{0,\pm}$, $\eta$,... channels. It can be viewed as an intrinsic
richness for mesons channels to allow for such flavor separation.

In summary, it should be clear that a full experimental program aiming
at the extraction of the individual GPDs is a broad program which requires 
the study of several isospin channels and several observables, each 
having its own characteristics. Only a \underline{global} study
and fit to all this information may allow an actual extraction of the GPDs.

\subsection{A dedicated facility}

An exploratory study of the GPDs can currently be envisaged at the 
JLab ($E_e$=6 GeV),
HERMES ($E_e$=27 GeV) and COMPASS ($E_\mu$=100-200 GeV) facilities
in a very complementary fashion, each having its own ``advantages" and ``disadvantages". The considerations which are relevant for this 
``exclusive" physics are~:
\begin{itemize}
\item Kinematical range : it is desirable to span a domain in $Q^2$ and
$x_B$ as large as possible, in particular to test scaling as mentionned
in section~\ref{sec1},
\item Luminosity : cross sections fall sharply with $Q^2$ and one has
to measure small cross sections,
\item Resolution : it is necessary to cleanly identify \underline{exclusive}
reactions. This can be achieved either by a good resolution with the
missing mass technique or by detecting \underline{all} the particles
of the final states and thus overdetermining the kinematics
of the reaction. 
\end{itemize}

Also, a large acceptance detector is desirable as $t$ and $\Phi$ 
(for asymmetries, studies of decay angular distributions,...)
coverages are needed and, more generally, the aim is, as emphasized
previously, to measure several channels and kinematic variables
simultaneously.

COMPASS, expected to start taking data in 2001, with a 100 to 200 GeV beam
has the clear advantage that it is the only facility
allowing to reach small $x_B$ (i.e. $\xi$) at sufficiently large $Q^2$. However,
it suffers from a relatively low luminosity ($\approx 10^{32} cm^{-2}s^{-1}$)
and relatively poor resolution to rely on the missing mass technique
in order to identify an exclusive reaction (there is a project of overcoming 
this latter point by 
adding a recoil detector which would overconstrain the kinematics of the 
reaction~\cite{nicole}). 

HERMES suffers basically from the same issues~:
relatively low luminosity ($\approx 10^{32-33} cm^{-2}s^{-1}$) and resolution not
fine enough to fully select exclusive final states, where, for instance, a 
typical missing mass resolution of the order of 300 MeV allows 
the contamination of additional pions into a sample of exclusive events. 
Here, however, a recoil detector is already under construction 
which should be operationnal soon and will overcome this issue~\cite{kaiser}. HERMES, which 
has been running since 1996, has the merit of being the first facility to 
have measured some experimental observables directly relevant to this physics 
($\rho^0_L$ cross sections, DVCS beam asymmetry, exclusive $\pi^+$ target asymmetry) as will be discussed in the next section.

JLab (with 6 GeV maximum beam energy in its current running configuration) has 
the highest luminosity ($\approx 10^{34} cm^{-2}s^{-1}$ for the Hall B large acceptance spectrometer 
in order to compare fairly with the other two facilities) and very good 
resolution (a typical missing mass resolution is less than 100 MeV. This 
good resolution is of course highly correlated with
the relatively low energy of the beam). The main drawback of JLab at 6 GeV
is obviously the limited kinematical range (at $x_B$=.3, W $>$ 2 GeV,
one cannot exceed $Q^2$=3.5 GeV$^2$ for instance). 

In spite of all the first ``breakthrough" measurements related to the
GPD physics that are currently being carried out at these facilities,
all these considerations clearly call for a dedicated machine which 
would combine a high luminosity ($\approx 10^{35-36} cm^{-2}s^{-1}$ desirable)
and a high energy ($\approx$ 30 GeV) beam with a good resolution detector 
(a few tens of MeV for a typical missing mass resolution). 
The ELFE~\cite{elfe} and JLab upgrade~\cite{cebaf11} (with a 11 GeV beam 
energy) projects would be quite well suited for such a physics program 
devoted to the systematic study of exclusive reactions and the GPDs.

\section{FIRST EXPERIMENTAL EVIDENCES}
\label{sec3}

\begin{figure}[htb]
\begin{minipage}[t]{80mm}
\epsfxsize=7. cm
\epsfysize=9. cm
\epsffile{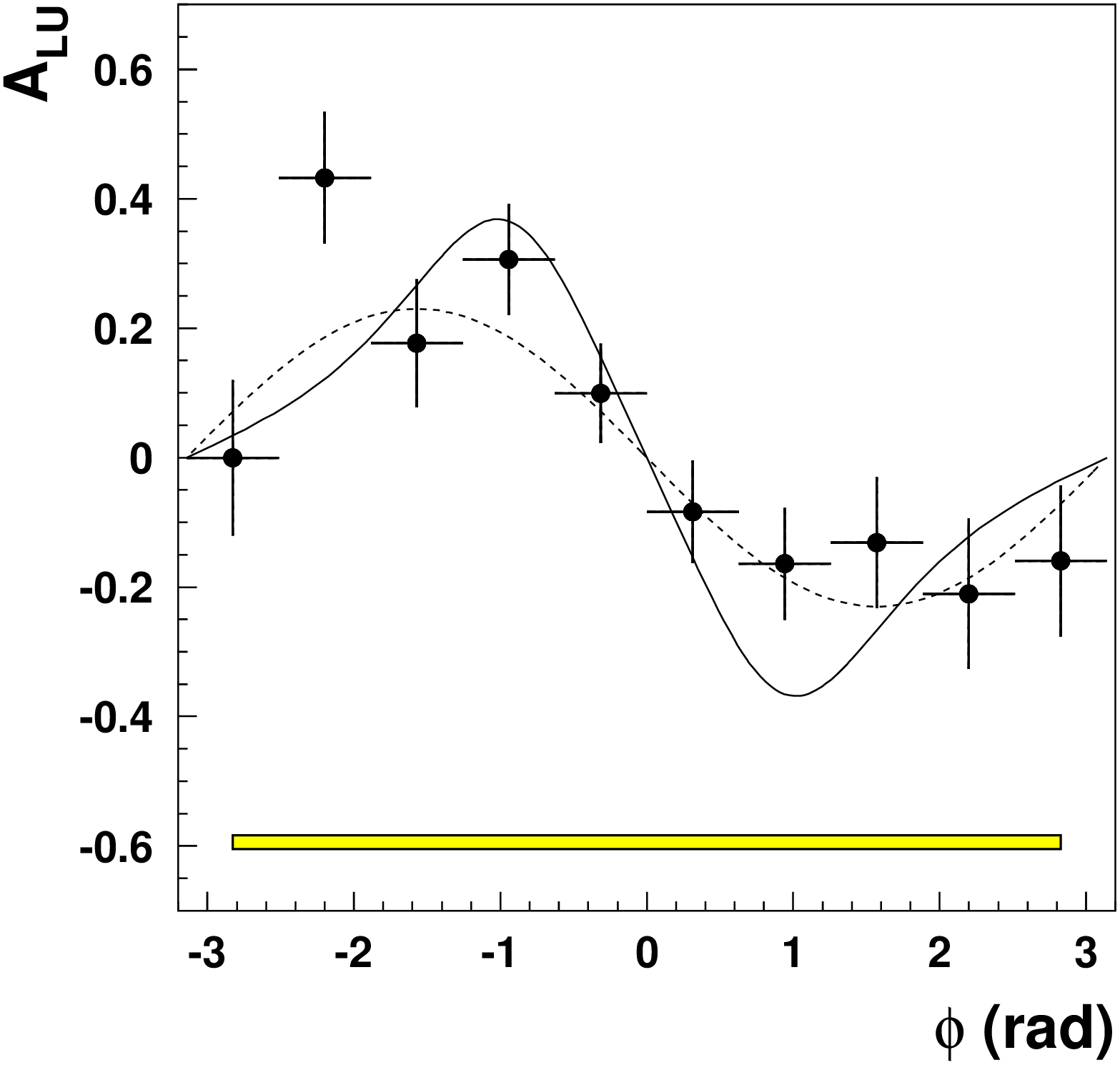}
\vspace{-3.2cm}
\caption{The DVCS beam asymmetry as a function of the azimuthal
angle $\Phi$ as measured by HERMES~\cite{dvcshermes}. Average kinematics 
is~: $<x>$=.11, $<Q^2>$=2.6 GeV$^2$ and $<-t>$=.27 GeV$^2$. The dashed curve
is a sin$\Phi$ fit whereas the solid curve is the theoretical GPD calculation
of Ref.~\cite{kivel}.}
\label{fig:hermes_dvcs}
\end{minipage}
\hspace{\fill}
\begin{minipage}[t]{75mm}
\epsfxsize=7. cm
\epsfysize=6.45 cm
\epsffile{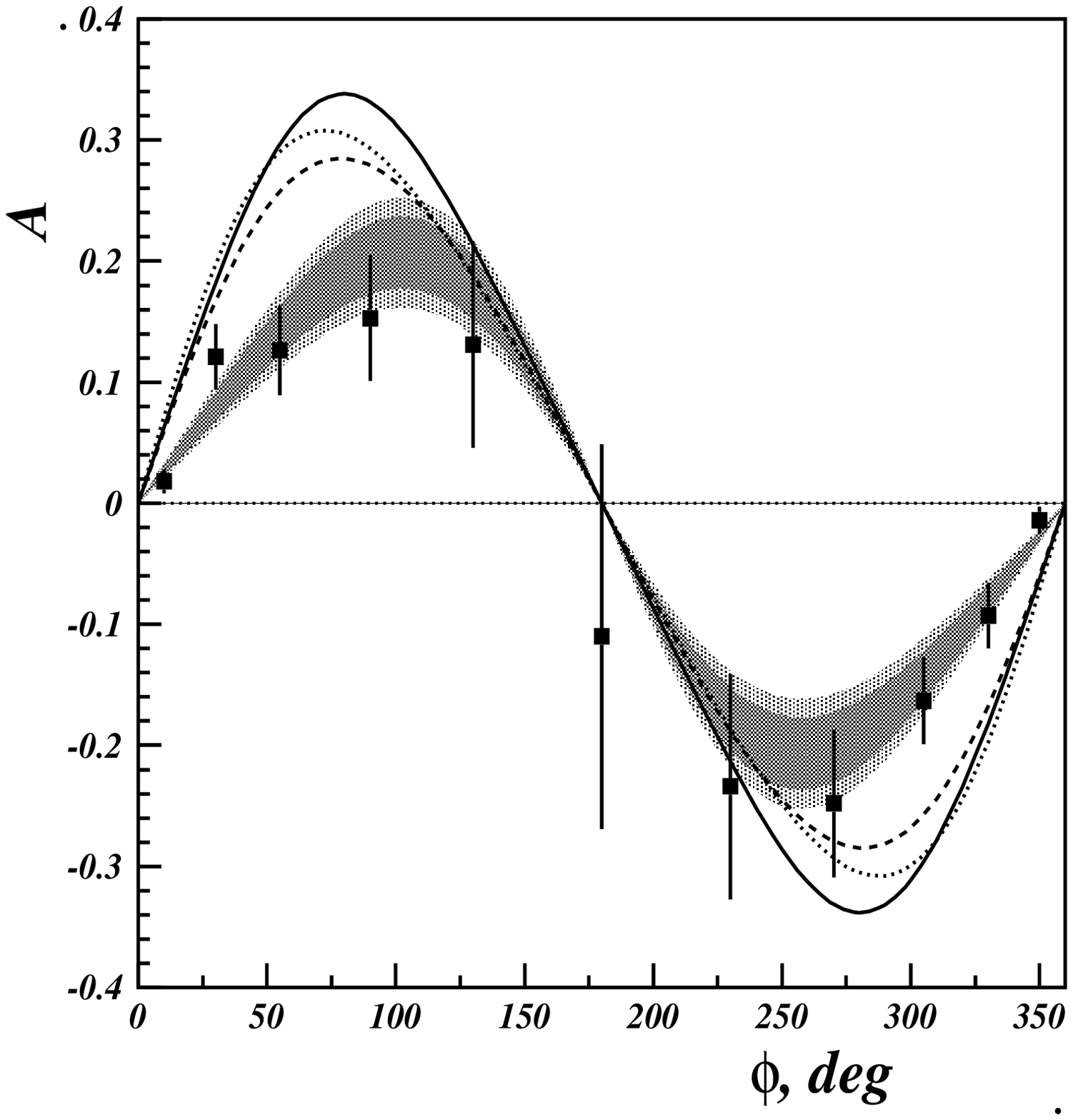}
\vspace{-3.2cm}
\caption{The DVCS beam asymmetry as a function of the azimuthal
angle $\Phi$ as measured by CLAS~\cite{dvcsclas}. Average
kinematics is~: $<x>$=.19, $<Q^2>$=1.25 GeV$^2$ and $<-t>$=.19 GeV$^2$.
The shaded regions are error ranges to sin$\Phi$ and sin$2\Phi$ fits.
Calculations are~: leading twist {\it without} $\xi$ dependence~\cite{vcsrev,marcprl} (dashed curve), 
leading twist {\it with} $\xi$ dependence~\cite{vcsrev,marcprl}
(dotted curve) and leading twist + twist-3~\cite{kivel} (solid curve).}
\label{fig:clas_dvcs}
\end{minipage}
\end{figure}
 
In this section, we review the first existing experimental data 
related to GPDs interpretation. Only these past 2 years, have been released
by the CLAS and HERMES collaborations experimental data precise
enough in the relevant kinematical regime. In this paper, we choose to focus
on the valence quark region, i.e. W $<$ 10 GeV, where the quark 
exchange mechanism of Fig.~(\ref{fig:handbags}) dominates. However,
it is to be mentionned that DVCS~\cite{fav} and vector mesons~\cite{adl} 
cross sections at low $x_B$ have also been measured by the H1 and ZEUS 
collaborations which lend themselves to GPD interpretation through 
``gluon exchange"-type processes~\cite{freund}.

Clearly, the statistics of all these 
measurements are still not high enough to allow for a fine
binning in the kinematical variables and therefore a precise 
test of GPD models. Nevertheless, they are very encouraging in the 
sense that the observed signals, although integrated over quite
wide kinematical ranges, are generally compatible (in magnitude 
and in ``shape") with theoretical calculations. It is to be noted by the way
that basically all of the calculations accompanying the figures
in the following were indeed {\it predictions} as they were published before 
the experimental results.

The experimental observables to be discussed are the single spin beam 
asymmetry (SSA) in DVCS (measured by HERMES and CLAS) and the longitudinal
cross section of $\rho^0$ electroproduction (measured by HERMES).
Fig.~(\ref{fig:hermes_dvcs}) shows the first measurement of the
SSA for DVCS by HERMES with a 27 GeV positron beam. This asymmetry
arises from the interference of the ``pure" DVCS process (where the outgoing
photon is emitted by the nucleon) and the Bethe-Heitler (BH) process (where
the outgoing photon is radiated by the incoming or scattered lepton).
The two processes are indistinguishable experimentally and interfer.
The BH process being purely real and exactly calculable in QED,
one has therefore access, through the difference of cross sections
for different beam helicities 
which is sensitive to the imaginary part of the amplitude,
to some {\it linear} combination of the GPDs at the kinematical point ($x=\xi,\xi,t$) as mentionned in the previous subsection.

The beam asymmetry, which is this latter difference of cross sections
divided by their sum, is more straightforward to access experimentally as 
normalization and systematics issues cancel, at first order, in the ratio.
For this asymmetry, a shape close to sin$\Phi$ (not an exact sin$\Phi$ 
shape as higher twists and the Bethe Heitler have some more complex
$\Phi$ dependence) is expected, where
$\Phi$ is the standard angle between the leptonic and the hadronic plane. 
At HERMES, the average kinematics 
is $<x>$=.11, $<Q^2>$=2.6 GeV$^2$ and $<-t>$=.27 GeV$^2$ for which an 
amplitude of .23 for the sin$\Phi$ moment is extracted from the 
fit~\cite{dvcshermes}. 
The discrepancy between the theoretical prediction and the data on 
Fig.~(\ref{fig:hermes_dvcs}) can certainly be attributed on the one hand to 
the large kinematical range over which the experimental data have been 
integrated and where the model can vary significantly and, on the other hand, 
to higher twists corrections not calculated (so far, only twist-3 are under 
theoretical control for the handbag DVCS process -see Ref.~\cite{diehl}-, the leading twist being twist-2).

Also, the DVCS reaction at HERMES is identified by detecting the scattered
lepton (\underline{positron}) and the outgoing photon from which the missing 
mass of the non-detected proton is calculated. Due to the limited resolution of 
the HERMES detector, the selected peak around the proton mass is 
$-1.5<M_x<1.7$ GeV which means that contributions to this asymmetry 
from nucleon resonant states as well cannot be excluded. Let's recall
that a recoil detector aiming at the detection of the recoil proton 
is projected to be installed at HERMES by 2003~\cite{kaiser}; this will 
then allow to unambiguously sign the exclusivity of the reaction at HERMES.

This same observable has been measured at JLab with a 4.2 GeV 
\underline{electron} beam with the 4$\pi$ CLAS detector~\cite{dvcsclas}. 
Due to the lower
beam energy compared to HERMES, the kinematical range accessed at JLab is different~: $<x>$=.19, $<Q^2>$=1.25 GeV$^2$ and $<-t>$=.19 GeV$^2$. 
In this case, the DVCS reaction was identified by detecting, besides 
the scattered lepton, the recoil proton and then calculating
the missing mass of the photon (due to the geometry of the CLAS
detector, the outgoing photon which is emitted at forward angles
escapes detection). The contamination by $ep\to ep\pi^0$ events
can be estimated and subtracted bin per bin, resulting in a 
rather clean signature of the exclusivity of the reaction.

Figure~(\ref{fig:clas_dvcs}) shows the CLAS measured asymmetry
along with theoretical calculations (predictions) which are in fair agreement
(the different sign of the CLAS SSA relative to HERMES is due to the
use of electron beams in the former case compared to positron beams in
the latter). 
Again, discrepancies can be assigned to the fact that the theory is 
calculated at a single well-defined kinematical point whereas data has 
been integrated over several variables and wide ranges. Furthermore,
Next to Leading Order as well twist-4 corrections which may be important 
at these rather low $Q^2$ values, still need to be quantified.

\begin{figure}[htb]
\begin{minipage}[t]{80mm}
\epsfxsize=11 cm
\epsfysize=6. cm
\centerline{\epsffile{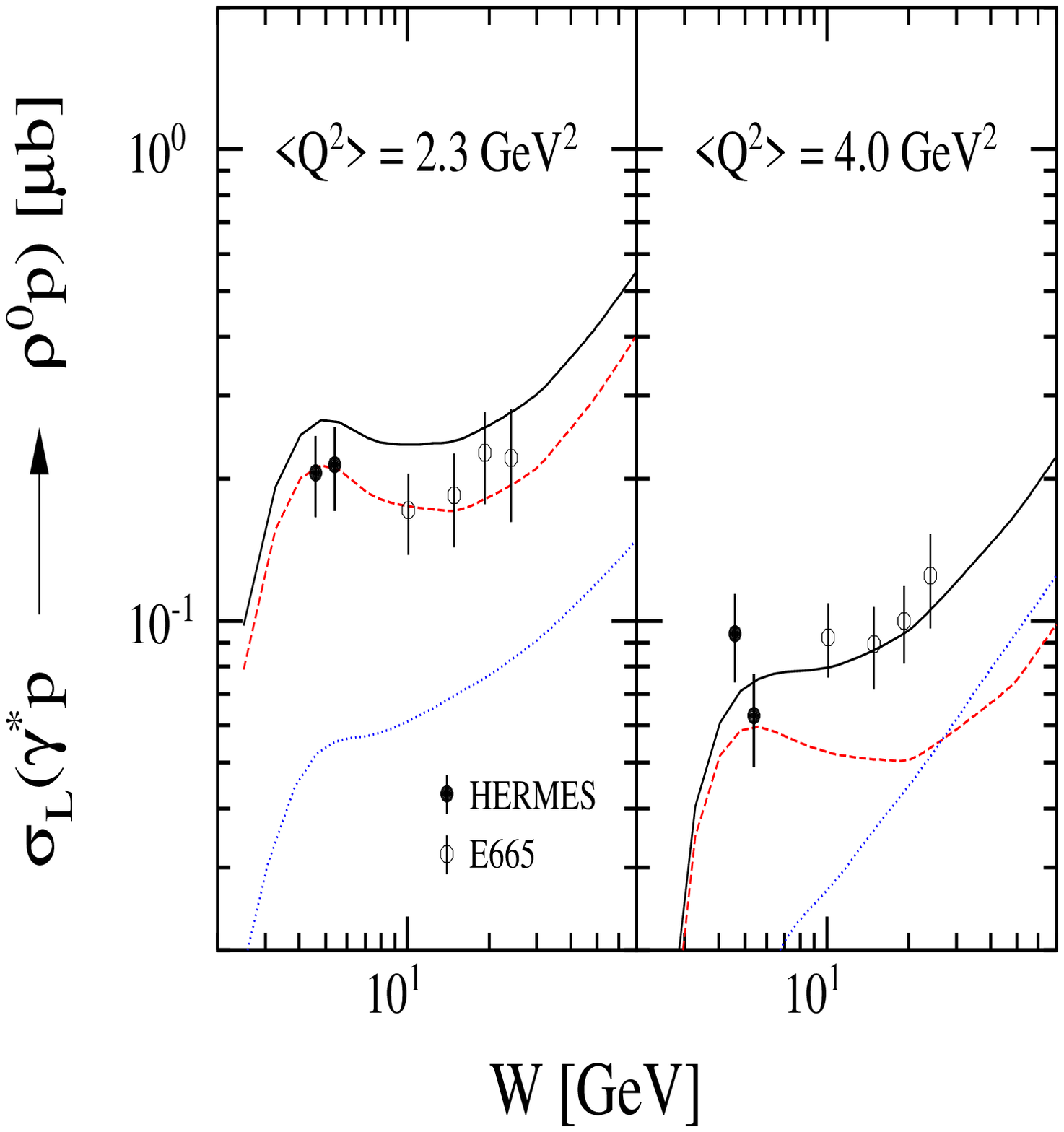}}
\vspace{-.9cm}
\caption{The longitudinal $\rho^0$ electroproduction cross section
in the intermediate $W$ range. The dashed curve represents the
quark exchange process calculated in the GPD framework whereas the dotted 
line represents the 2-gluon exchange process. The solid line
is the incoherent sum of the two mechanisms. Calculations are from Refs.~\cite{marcprl}. Figure is taken from Ref.~\cite{gerard}.}
\label{fig:rhoxsec}
\end{minipage}
\hspace{\fill}
\begin{minipage}[t]{75mm}
\epsfxsize=11 cm
\epsfysize=7.3 cm
\centerline{\epsffile{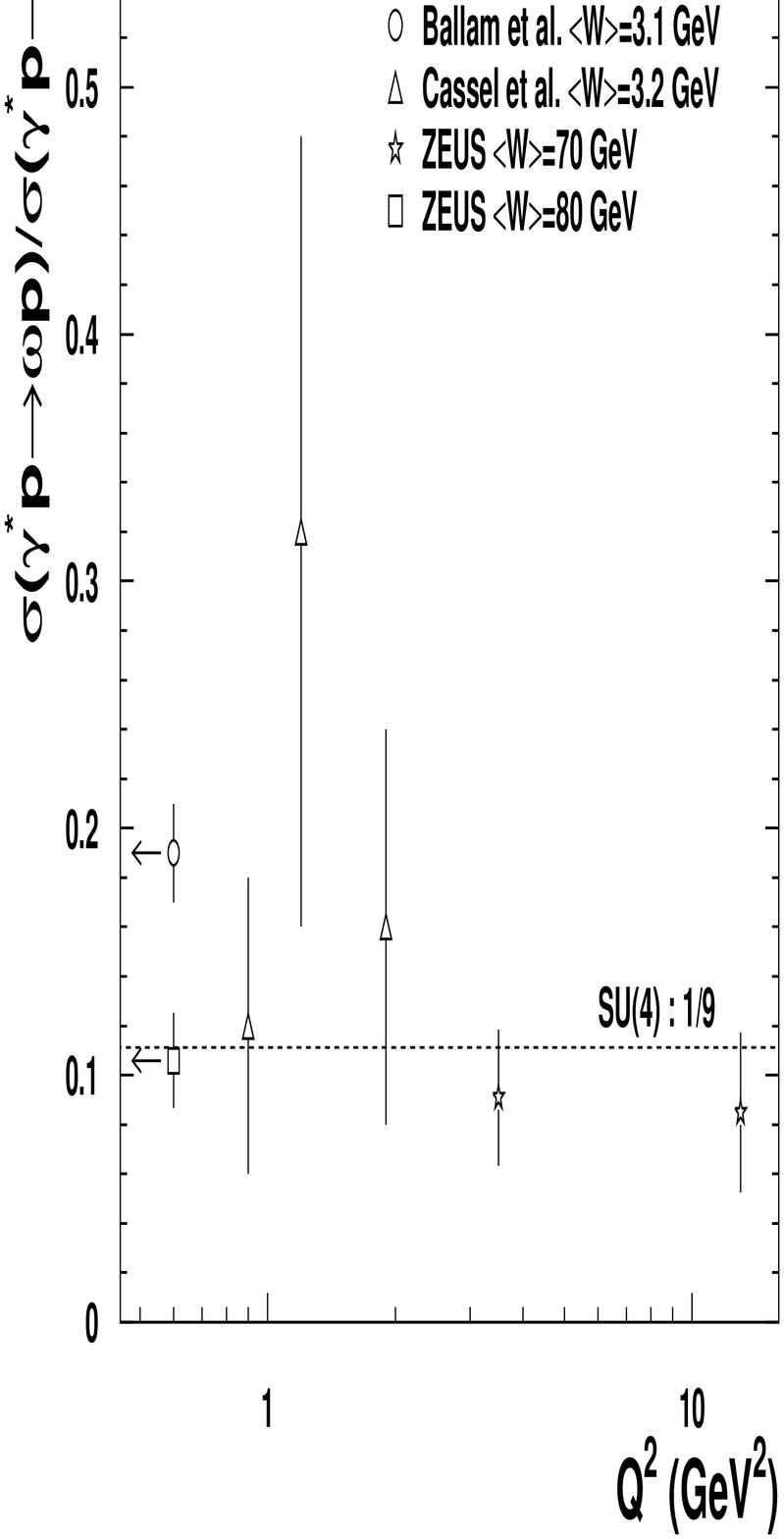}}
\vspace{-.9cm}
\caption{World data for the $\rho/\omega$ ratio as a function of $Q^2$.
The figure has been taken and adapted from~\cite{tytgat}.}
\label{fig:ratio}
\end{minipage}
\end{figure}

In the meson sector, the vector meson channel is the most accessible
as it allows rather simply to separate the longitudinal from the
transverse part of the cross section through its decay angular
distribution (we recall, as mentionned in section~\ref{sec1}, that
only for the longitudinal part of the cross section is the
factorization theorem valid at this stage and allows to make interpretations
in terms of GPDs). So far, only the $\rho^0$ channel has yielded 
sufficient statistics, due to its relatively high cross section, 
to isolate $\sigma_L$. Figure~(\ref{fig:rhoxsec}) shows the two HERMES
points~\cite{gerard} along with the GPD theoretical calculations (predictions). For vector mesons, two mechanisms contribute in two different kinematical regimes~: 
at low W (i.e. large $x_B$), 2-quark exchange ; at high W (i.e. low $x_B$), 
2-gluon exchange. The quark exchange process can be identified and 
calculated with the handbag diagrams of figure~\ref{fig:handbags}~\cite{vcsrev,marcprl}. 
For meson production, due to presence of the ``extra" gluon exchange
compared to DVCS, large corrections are expected to the leading order. 
These corrections can be modelled taking into account $k_\perp$ degrees of freedom~\cite{vcsrev,marcprl}.
At HERMES kinematics, this correction factor is found to be about 3 and allows
to predict the magnitude of the cross section.

One way to get rid of such model dependency in the corrections is to look at
\underline{ratios} of cross sections. Indeed, as pointed out by Ref.~\cite{Collins97,eides}, 
these correction factors are expected to factorize and therefore cancel in ratio.
One speaks of ``precocious scaling". The HERMES collaboration is about to
release the measurement of the $\omega$ cross section in the same
kinematical range as the $\rho^0$ cross section, it will be
very interesting to compare the $\omega\over\rho^0$ ratio to the
theoretical prediction of the GPD formalism which yields $\approx$ 1/5~\cite{vcsrev,marcprl},
this number, quite model independent, arising basically from the ratio of 
the u and d quark distributions weighted by known isopin factors.
This has to be compared to the well-known SU(3) 1/9 prediction in the 
low $x_B$ domain. A $W$ (or equivalently $x_B$) dependence is therefore
expected for this ratio. This seems to be already observed with the
current world data, see Fig.~(\ref{fig:ratio}), where one can already 
distinguish a trend -in spite of quite large error bars-  where the low 
$W$ data are close to $\approx$ 1/5 whereas the large $W$ data are closer 
to 1/9. The preliminary HERMES results tend to confirm this tendency~\cite{tytgat}.

Similarly, $\pi^+\over\pi^0$, $\pi^0\over\eta$, $\rho^+\over\rho^0$,
etc... ratios deserve to be measured as they can be directly compared
to leading order and leading twist model independent theoretical predictions
in the GPD framework.

\section{Conclusion}

In conclusion, we believe that the GPDs open a broad new area of physics, 
by providing a context for understanding \underline{exclusive} reactions 
in the valence region (where the quark exchange mechanism dominates) at large $Q^2$.
By ``constraining" the final state of the DIS reaction, instead of
summing over all final states, one accesses some more fundamental structure
functions of the nucleon. These functions provide
a unifying link between a whole class of various reactions 
(elastic and inelastic) and fundamental quantities as diverse as
form factors and parton distributions. They allow to access new information
on the structure of the nucleon, for instance quark's orbital momentum
and, more generally, correlations between quarks.

A full study aiming at the extraction of these GPDs from experimental
data requires a new dedicated facility providing high energy and high
luminosity lepton beams, equipped with large acceptance and high resolution
detectors. First experimental exploratory results from the HERMES and CLAS 
collaboration provide some evidence that the manifestations of the handbag
mechanisms are already observed. This is encouraging and paves the
way for a future very rich harvest of hadronic physics and motivates
the development of new dedicated projects and facilities.

\end{document}